\documentclass[journal]{IEEEtran}
\IEEEoverridecommandlockouts 
\usepackage{cite}
\usepackage{xspace}
\usepackage{caption, subcaption}
\usepackage{mathtools,amssymb}
\usepackage{bm}
\usepackage{booktabs}
\usepackage[linesnumbered,ruled,lined]{algorithm2e}
\usepackage{url}  
\usepackage{amsthm}
\usepackage{multirow}
\usepackage{color}
\usepackage{endnotes}
\usepackage[subnum]{cases}
\usepackage[nocomma]{optidef}
\usepackage{graphicx}

\begin{document}
\title{Data Service Maximization in Space-Air-Ground Integrated 6G Networks
\thanks{}}
%
\author{Nway~Nway~Ei,~Kitae Kim,~Yan~Kyaw~Tun,~\IEEEmembership{Member,~IEEE},\\~Zhu~Han,~\IEEEmembership{Fellow,~IEEE,~}~and~Choong~Seon~Hong,~\IEEEmembership{Fellow,~IEEE}
\thanks{Nway Nway Ei, Kitae Kim and Choong Seon Hong  are with the Department of Computer Science and Engineering, Kyung Hee University,  Yongin-si, Gyeonggi-do, South Korea, 436-701, Rep. of Korea, e-mail: {\{nwayei, glideslope, cshong\}@khu.ac.kr}.}
\thanks{Zhu Han is with the Electrical and Computer Engineering Department, University of Houston, Houston, TX 77004, and the Department of Computer Science and Engineering, Kyung Hee University, Yongin-si, Gyeonggi-do, South Korea, 436-701,  Rep. of Korea, email{\{hanzhu22\}@gmail.com}.}
\thanks{Yan Kyaw Tun is with the Department of Electronic Systems, Aalborg University, A. C. Meyers Vænge 15, 2450 København, e-mail: ykt@es.aau.dk.}
}

\maketitle

\begin{abstract}
Integrating terrestrial and non-terrestrial networks has emerged as a promising paradigm to fulfill the constantly growing demand for connectivity, low transmission delay, and quality of services (QoS). This integration brings together the strengths of the reliability of terrestrial networks, broad coverage and service continuity of non-terrestrial networks like low earth orbit satellites (LEOSats), etc. In this work, we study a data service maximization problem in space-air-ground integrated network (SAGIN) where the ground base stations (GBSs) and LEOSats cooperatively serve the coexisting aerial users (AUs) and ground users (GUs). Then, by considering the spectrum scarcity, interference, and QoS requirements of the users, we jointly optimize the user association, AU's trajectory, and power allocation. To tackle the formulated mixed-integer non-convex problem, we disintegrate it into two subproblems: 1) user association problem and 2) trajectory and power allocation problem. We formulate the user association problem as a binary integer programming problem and solve it by using the Gurobi optimizer. Meanwhile, the trajectory and power allocation problem is solved by the deep deterministic policy gradient (DDPG) method to cope with the problem's non-convexity and dynamic network environments. Then, the two subproblems are alternately solved by the proposed block coordinate descent algorithm. By comparing with the baselines in the existing literature, extensive simulations are conducted to evaluate the performance of the proposed framework.
\end{abstract}

\vspace{-0.3cm}
\section{Introduction}
In addition to reliable and high-speed data services, the sixth generation (6G) network is anticipated to offer end-user devices ubiquitous connectivity by integrating conventional terrestrial networks and non-terrestrial networks such as low earth orbit satellites (LEOSats). According to the recent report of SpaceX, the LEO satellites with direct-to-cell capabilities are expected to deploy more for the broader network coverage \cite{starlink}. With the effective cooperation of terrestrial and satellite network operators, all users (aerial and ground users) can achieve their QoS requirements.  

Despite their prominent advantages, the space-air-ground integrated network (SAGIN) faces several challenges, such as dynamic resource management, load balancing, and interference management between terrestrial and non-terrestrial networks. There are existing works that address the interference management problem due to the spectrum sharing between terrestrial and non-terrestrial networks \cite{zhu2017non, fang2022noma, lei2024spatial, lee2023interference}. For instance, the authors in \cite{zhu2017non} proposed an integrated terrestrial and satellite network where a non-orthogonal multiple access (NOMA) scheme is adopted to enhance the capacity of the ground users. Similarly, the work in \cite{fang2022noma} investigated the interference-aware sum rate maximization problem in a space-air-ground network. Considering the full frequency reuse among the LEOSats, the authors in \cite{lei2024spatial} proposed a dynamic beamforming and user scheduling problem for uneven traffic demands. To alleviate the radio resource shortage problem in the SAGIN network, the authors in \cite{lee2023interference} proposed an efficient reverse spectrum sharing framework for the overall system capacity maximization. Moreover, the authors in \cite{chen2022multi} proposed a collaborated satellite and terrestrial network where the data traffic of terrestrial network is offloaded to the satellite network. By deploying the unmanned aerial vehicles (UAVs) as the relay nodes, the work in \cite{choi2024latency} proposed a learning-based resource allocation algorithm. 

The aforementioned works \cite{zhu2017non, fang2022noma, lei2024spatial, lee2023interference, chen2022multi, choi2024latency} only focus on interference-aware resource management for the ground users (GUs) and did not consider the existence of aerial users (AUs). Therefore, in this work, we study a SAGIN network where the ground base stations (GBSs) and LEOSats collaboratively serve the coexisted AUs and GUs in the downlink by sharing the same spectrum resources. However, due to the mobility of LEOSats and AUs, it is challenging to allocate the radio resources to all the users (both aerial and ground users) in real time. Moreover, the trajectories of AUs should be properly optimized to mitigate the severe interference in the network. It is also challenging for users to determine the association between GBSs and LEOSats to avoid high interference and maintain good network connectivity. Therefore, the main contributions of this paper are summarized as follows:
\begin{itemize}
    \item Firstly, we propose a novel SAGIN for 6G where, the GBSs and LEOSats cooperatively serve the coexisting AUs and GUs in the downlink.
    \item We then formulate the data service maximization problem to jointly optimize the user association, AUs' trajectories, and power allocation of GBSs and LEOSats. To address the formulated non-convex problem, we decompose it into two sub-problems; 1) association problem and 2) trajectory and power allocation problem. These sub-problems are solved in a block coordinate descent manner to obtain the solution of the proposed optimization problem.
    \item Finally, we conduct the extensive simulations to verify the performance of our proposed algorithm. The evaluation results indicate that the data service achieved by our proposed algorithm surpasses that of the existing baseline methods. Statistically, our proposed algorithm provides $51.6$ \% improvement over random association (RA), $19.2$ \% improvement over double deep Q-learning-based trajectory optimization (T-DDQN), and $10.4$ \% improvement over equal power allocation (EPA).  
\end{itemize}
\section{System Model and Problem Formulation} 
\begin{figure}[t!]
	\centering
	\includegraphics[width=0.85\linewidth, height=4.8cm]{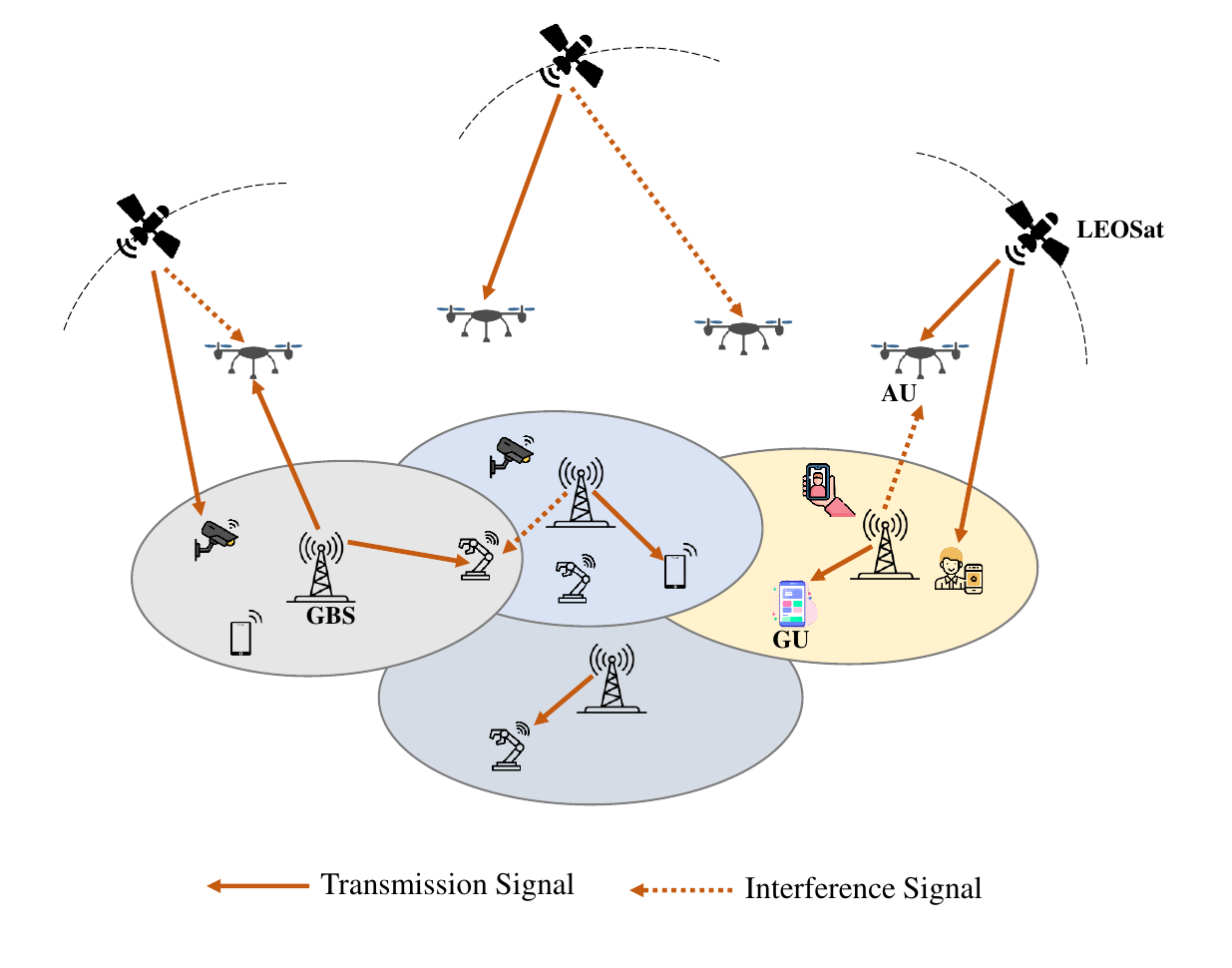}
	\caption{Space-air-ground integrated 6G network.}
	\label{fig:system_model}
\end{figure}

We propose a SAGIN as depicted in Fig. \ref{fig:system_model}, where the GBSs and LEOSats cooperatively serve the AUs and GUs for the downlink communication. The set of GBSs and LEOSats are defined as $\mathcal{N}=\{1,2,...,N\}$ and $\mathcal{S}=\{1,2,...,S\}$, respectively. For simplicity, the GBSs and LEOSats are denoted as base stations (BSs) defined by a set $\mathcal{B} = \mathcal{N} \cup \mathcal{S}=\{1,2,...,B\}$. In the considered area, there is a set of AUs $\mathcal{I}=\{1,2,...,I\}$ and a set of GUs $\mathcal{J}=\{1,2,...,J\}$. Similarly, the set of all users is defined as $\mathcal{U} = \mathcal{I} \cup \mathcal{J}=\{1,2,...,U\}$, where each user can access either GBS or LEOSat for data services such as high-precision map downloading, infotainment services, or other mission-related data services. The set of subchannels that each BS can occupy in the proposed SAGIN is defined as $\mathcal{K} = \{1,2, ...,K\}$. Due to the spectrum scarcity, the GBSs and LEOSats are supposed to share the same frequency band to serve their associated users. However, the orthogonal frequency division multiple access (OFDMA) framework is implemented at each BS while serving its associated users. Therefore, there is no intra-cell interference; the users will only experience the inter-cell interference. Furthermore, the time interval is divided into a number of time slots defined by a set $\mathcal{T} = \{1,2, ...,T\}$. At the beginning of each time slot, the information of users including their positions and demand is sent to the gateway with the help of GBSs and LEOSats. Then, the optimized decisions are determined at the gateway connected to the core network and informs back to the GBSs and LEOSats.

Considering the free space path loss, shadow fading and rain attenuation, the channel gain between LEOSat $s$ and user $u \in \mathcal{U}$ is given by \cite{tsai2024distributionally}; $\nu_{u,s}[t] = 10^{-L_{u,s}[t]/10}G_s^tG_u^r$, where $L_{u,s}[t]$ is the path loss and $G_s^t$ and $G_u^r$ are the antenna gains of LEOSat $s$ and user $u$, respectively. For the link between AU and GBS, we adopt Rician channel model and the channel gain between AU $i$ and GBS $n$ is given as $\nu_{i,n}[t] = \frac{\nu_0}{(\lVert z_i[t] - q_n \rVert^2 + (h_i[t] - h_n)^2)}$; where $z_i[t] = (x_i[t], y_i[t])$ and $q_n = (x_n, y_n)$ are the coordinates of AU $i$ and GBS $n$. $h_i[t]$ and $h_n$ are the altitudes of AU $i$ and GBS $n$, respectively. $\nu_0$ is the channel gain at reference distance of $1$ m. However, the channel gain between GU $j$ and GBS $n$ at time slot $t$ can be calculated by \cite{pervez2021joint}; $\nu_{j,n}[t] = \frac{\nu_0 \delta}{(\lVert q_j - q_n \rVert^2 + h_n^2)}$  
where $\delta$ is the random variable that accounts for small-scale Rayleigh fading.

Since the subchannels are shared among the BSs to serve the users, the signal to interference plus nois ratio (SINR) of user $u$ when it associates with BS $b$ in time slot $t$ is given as
\begin{equation}
    \gamma_{u,b}^k[t] = \frac{|\nu_{u,b}[t]|^2p_b^k[t]}{I_{u,b}^k[t] + \sigma^2},
\end{equation}
where $I_{u,b}^k[t] = \sum_{\substack{b' \in \mathcal{B},\\ b' \ne b}}|\nu_{u,b'}[t]|^2p_{b'}^k[t]$ is the inter-cell interference from other BSs and $p_b^k[t]$ is the transmit power of BS $b$ on subchannel $k$. $\sigma^2$ is the noise. Therefore, the achievable data rate of user $u$ when it associates with BS $b$ on subchannel $k$ in time slot $t$ can be given as
\begin{equation}
   R_{u,b}^k[t] = w_k\log_2\big(1 + \gamma_{u,b}^k[t]\big),
\end{equation}
where $w_k$ is the channel bandwidth.

For user association, assuming that each user can only be assigned with one subchannel in each time slot, the association variable is defined as $\alpha_{u,b}^k[t] \in \{0,1\}$; its value is $1$ if user $u$ is associated with BS $b$ on subchannel $k$ in time slot $t$; $0$, otherwise.
Finally, the data service of the users is defined as the amount of data downloaded in time slot $t$, which is given as
\begin{equation}
    D^{\mathrm{tot}}[t] = \tau\sum_{b=1}^{B}\sum_{u=1}^{U}\sum_{k=1}^{K}\alpha_{u,b}^k[t]R_{u,b}^k[t],
\end{equation}
where $\tau$ is the duration of time slot $t$. 

Given the system model, our objective is to maximize the data service of all users by jointly optimizing the user association, AU's trajectory, and BS's power allocation. Taking into account the limited network resources and QoS requirements of the users, the data service maximization problem is formulated as follows:
\vspace{-0.25cm}
\begin{maxi!}[2]                 
		{\boldsymbol{\alpha}, \boldsymbol{z}, \boldsymbol{p}}                               
		{\sum_{t=1}^{T} D^{\mathrm{tot}}[t]}{\label{opt:P}}{\textbf{P:}} 
		\addConstraint{\alpha_{u,b}^k[t] \in \{0,1\}, \forall u \in \mathcal{U} , b \in \mathcal{B}, k \in \mathcal{K}, t \in \mathcal{T},}
        \addConstraint{\sum_{b=1}^{B}\sum_{k=1}^{K}\alpha_{u,b}^k[t] \le 1, \forall u \in \mathcal{U}, t \in \mathcal{T},}
		\addConstraint{\sum_{u=1}^{U}\alpha_{u,b}^k[t] \le U_{\textrm{max}}, \forall b \in \mathcal{B}, t \in \mathcal{T},}
		\addConstraint{p_b^k[t] > 0, \sum_{u=1}^{U}\alpha_{u,b}^k[t]p_b^k[t] \le P_{\textrm{max}}, \forall b \in \mathcal{B}, } \nonumber
        \addConstraint{\ \ \ \ \ \ \ \ \ \ \ \ \ \ \ \ \ \ \ \ \ \ \ \ \ \ \ \ \ \ \ \ \ \ \ \ \ \ \ \ \ \ \ t \in \mathcal{T},} 
\addConstraint{\sum_{b=1}^{B}\sum_{k=1}^{K}\alpha_{u,b}^k[t]\gamma_{u,b}^k[t] \ge \gamma_0, \forall u \in \mathcal{U}, t \in \mathcal{T},}
        \addConstraint{\lVert z_i[t] - z_i[t-1] \rVert \le V_{\textrm{max}}\tau, \forall i \in \mathcal{I},}
        \addConstraint{\lVert z_i[t] - z_{i'}[t] \rVert \ge d_{\textrm{min}}, \forall i,i' \in \mathcal{I}, i \neq i', t \in \mathcal{T},}
        \addConstraint{z_i[0] = z_0, z_i[T] = z_f, \forall i \in \mathcal{I},}
        \addConstraint{h_{\textrm{min}} \le h_i[t] \le h_{\textrm{max}}, \forall i \in \mathcal{I}, t \in \mathcal{T}},
  \vspace{-0.1cm}
\end{maxi!}
where (4b) and (4c) ensure that each user can only be allocated with one subchannel in each time slot. Furthermore, due to the limited available subchannels, the maximum number of users each BS can serve must not exceed $U_{\textrm{max}}$ as given in (4d). Then, according to (4e), the allocated power of each BS to its associated users must not exceed its maximum power. Constraint (4f) guarantees the QoS requirement of each user in time slot $t$. Finally, the trajectory constraints that reflect AU's maximum velocity, collision avoidance, and altitudes are given in (4g)-(4j). 
\vspace{-0.4cm}
\section{Proposed Solution}
This section presents our solution approach to solve the formulated mixed-integer non-convex problem in \textbf{P}. To address the coupling of the binary association variables and continuous trajectory and power allocation variables, we decompose problem \textbf{P} into two tractable subproblems, namely, 1) association problem and 2) trajectory and power allocation problem. Then, the two subproblems are iteratively solved to obtain the solution of \textbf{P}.         
\subsubsection{Association Problem}
Given the trajectory of AUs and power allocation of BSs, the association problem can be rewritten as
\vspace{-0.2cm}
\begin{maxi!}[2]                 
		{\boldsymbol{\alpha}}                               
		{\sum_{t=1}^{T} D^{\mathrm{tot}}[t]}{\label{opt:P1}}{\textbf{P1:}} 
		\addConstraint{\text{(4b)}-\text{(4f)}.}
	\end{maxi!}
Since the problem \textbf{P1} is the binary integer programming problem, we use the Gurobi optimizer to obtain the association of users with BSs.  
\subsubsection{Trajectory and Power Allocation Problem}
With the association values obtained from subproblem \textbf{P1}, we reformulate the trajectory and power allocation problem as
\begin{maxi!}[2]                 
		{\boldsymbol{z}, \boldsymbol{p}}                               
		{\sum_{t=1}^{T} D^{\mathrm{tot}}[t]}{\label{opt:P2}}{\textbf{P2:}} 
		\addConstraint{\text{(4e)}-\text{(4j)}.}
\end{maxi!}
It is challenging to solve {\textbf{P2}} because of the coupling of the trajectory and power allocation variables. Moreover, the objective function in (4a) and constraints (4f) and (4g) are non-convex with respect to $\boldsymbol{z}$ and $\boldsymbol{p}$. Due to the mobility of the AUs and LEOSats, it is challenging and computationally highly complex to use conventional optimization methods to cope with the dynamic network environment and achieve the optimal solution of {\textbf{P2}}. Therefore, we propose a deep deterministic policy gradient (DDPG)-based trajectory and power allocation algorithm to solve {\textbf{P2}}. Generally, DDPG is a deep reinforcement learning algorithm that utilizes the policy gradient method to handle the continuous action spaces. We define state, action, and reward corresponding to \textbf{P2} as follows:
\begin{itemize}
    \item \textit{State:} The state space $s(t)$ of the system contains the association of users with BSs, the position of AUs, and the QoS requirements of all users in the current time slot.
    \item \textit{Action:} The action space consists of the trajectory of AUs and the power allocation of BSs. Here, we consider the horizontal flying velocity $v_i^h[t]$, directional angle $\theta_i[t]$, and the height $h_i[t]$ of AUs as their trajectories. Therefore, the action space is defined as $a(t) = \{v_i^h[t], \theta_i[t], h_i[t], p_b^k[t], \forall i \in \mathcal{I}, k \in \mathcal{K}, b \in \mathcal{B}\}$.
    \item \textit{Reward:} By taking into account the flying energy of AUs, we define the reward function for {\textbf{P2}} as
\end{itemize} 
\begin{algorithm}[t!]   
\caption{DDPG-Based Trajectory and Power Optimization Algorithm}
\label{alg:ddpg_table}
\scriptsize{
Initialize the parameters $\phi^\mu$ and $\phi^{\mu'}$ of the policy network $\mu(.)$ and target policy network $\mu'(.)$, and set $\phi^{\mu'} = \phi^\mu$; \\
Initialize the parameters $\phi^Q$ and $\phi^{Q'}$ of the Q-network $Q(.)$ and target Q-network $Q'(.)$, respectively, and set $\phi^{Q'} = \phi^Q$;  \\
\For{episode $= 1$ to $E$}{
    Initialize the random process $\Omega$ for the exploration of the action; \\
    Observe state $s(1)$; \\
    \For{$t \in \mathcal{T}$}{
    \While{$(4e)$-$(4j)$ are not met and until the maximum step has reached}{
         According to the current policy and exploration noise, select the action $a(t) = \mu\big(s(t)|\phi^\mu\big) + \Omega_t$; \\
        \If{The constraints are deviated}{
            Cancel the action and apply the penalty; 
        }
        Store the transition $(s(t), a(t), r'\big(s(t), a(t)\big), s(t+1))$ in experience replay buffer $B$; 
  }
    Random minibatch of $H$ transitions $(s_i, a_i, r_i, s_{i+1})$ from $B$ are sampled; \\
    Compute the target Q-value $y(i)=$ 
    \vspace{-0.12in}
               \[r'(i) + \gamma Q'\bigg(s(i+1), \mu'\big(s(i+1)|\phi^{\mu'}\big)|\phi^{Q'}\bigg); \] \\
    \vspace{-0.12in}
    Update the weights $\phi^Q$ of $Q(.)$ by minimizing the loss function 
    \vspace{-0.12in}
            \[
            L(\phi^Q) = \frac{1}{H}\sum_{i=1}^{H}\bigg(y(i)-Q\big(s(i), a(i)|\phi^Q\big)\bigg);
            \]    \\
    \vspace{-0.12in}
    Update the target Q-network parameters: $\phi^{Q'} = \eta\phi^Q + (1-\eta)\phi^{Q'}$; \\
    Update the weights $\phi^\mu$ of the policy network $\mu(.)$ by applying the gradient method as in \text{(11)}; \\
    Update the target policy network: $\phi^{\mu'}= \eta\phi^\mu + (1 - \eta)\phi^{\mu'}$; \\
    Obtain the optimal flying velocity, directional angle, and height of the AUs; \\
    Obtain the positions of AUs by 
    $x_i[t+1] = x_i[t] + \tau v_i^h[t]\cos{(\theta_i[t])}$, 
    $y_i[t+1] = y_i[t] + \tau v_i^h[t]\sin{(\theta_i[t])}$
    ;
    }
    Obtain $\boldsymbol{z}^*$ and $\boldsymbol{p}^*$.
}
}
\vspace{-0.1cm}
\end{algorithm}
\begin{equation}
    r(t) = D^{\mathrm{tot}}[t] - \beta\sum_{i=1}^{I}E_i^{\textrm{fly}}[t],  \label{reward_eq}
\end{equation}
where $\beta$ is the weight parameter, and the flying energy of AUs can be calculated by \cite{mei2019joint} and \cite{jiang2022green},
\begin{multline*}
    E_i^{\textrm{fly}}[t] = \tau \Biggr[A_0\bigg(1 + \frac{3(v_i^h[t])^2}{U_{\mathrm{tip}}^2}\bigg) + \frac{1}{2}\psi_0 \Tilde{r}\rho G(v_i^h[t])^3 \\ + A_1\bigg(\sqrt{1 + \frac{(v_i^h[t])^4}{4v_o^4}} - \frac{(v_i^h[t])^2}{2v_0^2}\bigg)^{\frac{1}{2}} + A_2v_i^v[t]\Biggr], 
\end{multline*}
where $A_0$ and $A_1$ are constants for blade profile power and induced power in the hovering state. $A_2$ is the ascending/descending power related constant and $U_{\mathrm{tip}}$ is the tip speed of rotor blade. $\psi_0$ and $\Tilde{r}$ are the fuselage drag ratio and rotor solidity, respectively. $\rho$ and $G$ are the air density and rotor disc area, respectively. $v_0$ is the mean rotor-induced velocity for hovering. $v_i^v[t]$ is the vertical flying velocity of AU $i$ that can be calculated as $v_i^v[t] = \frac{|h_i[t] - h_i[t-1]|}{\tau}$.  

To guarantee that all the constraints (4e)-(4j) are met, we modify the reward function in (\ref{reward_eq}) as follows:
\begin{equation}
    r'(t) = r(t) - a_1 - a_2 - a_3,  
\end{equation}
where $a_1$ and $a_2$ are penalty terms that reflect constraints (4e) and (4f). $a_3$ is the penalty term for trajectory constraints in (4g)-(4j). Then, the policy network generates the action value by learning the network parameter $\phi^{\mu}$ as follows:
\begin{equation}
    a(t) = \mu\big(s(t)|\phi^{\mu}\big).       
\end{equation}

After choosing the action based on the current state of the environment, the deep Q-network evaluates the selected action by using the Q-value as 
\begin{equation}
    Q\big(s(t), a(t)\big) = \mathop{\mathbb{E}}\big[r'\big(s(t), a(t)\big) + \gamma Q\big(s(t+1), a(t+1)\big)\big].    
\end{equation}
By implementing the target Q-network, the Q-network is trained to minimize the loss between them. The policy network parameter $\phi^{\mu}$ can be updated by the gradient method as 
\begin{equation}
\begin{split}
    \nabla_{\phi^\mu}J(\mu) & \approx 
    \frac{1}{H}\sum_{i}^{H}\bigg(\nabla_aQ\big(s, a|\phi^Q\big)|_{s=s(i), a=\mu\big(s(i)\big)} \\  & \nabla_{\phi^\mu}\mu\big(s|\phi^\mu\big)|_{s(i)}\bigg),
\end{split}    
\end{equation}
where $J(\mu)$ is the output of the policy network. The proposed DDPG-based trajectory and power allocation algorithm is given in Algorithm \ref{alg:ddpg_table}. 

To obtain the solution of \textbf{P}, we solve two subproblems in a block coordinate descent manner. The computation complexity of our proposed data service maximization problem is $\mathcal{O}\bigg(\hat{r}T\big(UB + E\sum_{\hat{h}=1}^{\hat{H}-1}\hat{\iota}_{\hat{h}}\hat{\iota}_{\hat{h}+1}\big)\bigg)$, where $\hat{H}$ is the number of hidden layers and $\hat{\iota}_{\hat{h}}$ is the number of neurons in each layer. $\hat{r}$ is the iteration numbers. 
\section{Numerical Results}     \label{sim_results}
In this section, we provide the numerical results to verify the performance of the proposed algorithm in the SAGIN. We consider the square area of $1.3$ km $\times$ $1.3$ km, with $4$ GBSs and $2$ LEOSats. The heights of the LEOSats are set as $[300, 700]$ km. The range of the values of $v_i^h[t]$, $\theta_i[t]$, and $h_i[t]$ are set as $[10, 20]$ m/s, $[\frac{-5\pi}{12}, \frac{5\pi}{12}]$, and $[200, 300]$ m, respectively. The coordinates of GUs are randomly generated in the considered area. Moreover, the maximum transmission power of BSs is set as $[30, 50]$ W and the subchannel bandwidth is set as $180$ kHz. Then, several baselines are implemented to compare with the proposed algorithm as follows: 
\begin{figure}[t!]
  \centering
  \begin{subfigure}[b]{0.48\columnwidth}
    \centering
    \includegraphics[width=\textwidth]{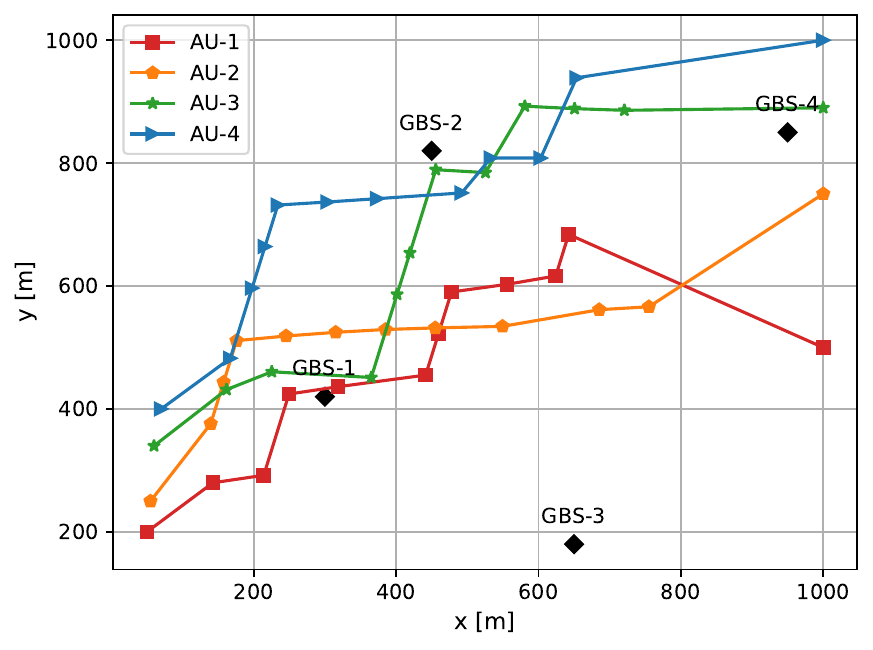}
    \caption{Proposed.}
    \label{fig:traj_proposed}
  \end{subfigure}
  \begin{subfigure}[b]{0.48\columnwidth}
    \centering
    \includegraphics[width=\textwidth]{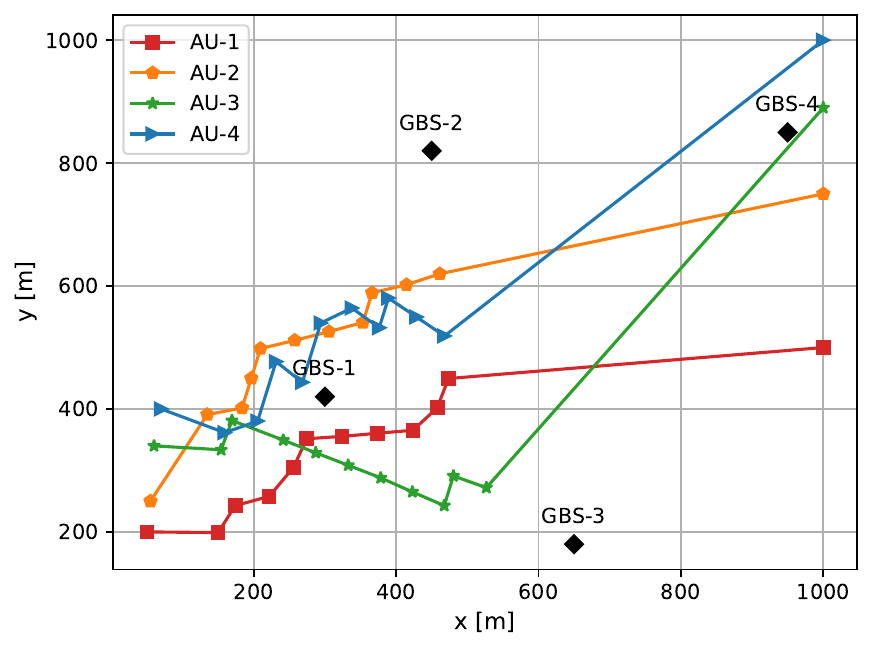}
    \caption{T-DDQN.}
    \label{fig:traj_DDQN}
  \end{subfigure}
  \caption{Optimal trajectories of AUs under different algorithms.}
  \label{fig:opt_traj}
\end{figure}
\begin{itemize}
        \item \emph{T-DDQN}: In this baseline, the user association and power allocation are addressed by our proposed algorithm while the DDQN algorithm solves the trajectory optimization problem.
        \item \emph{RA}: In this framework, we randomly associate the users with BSs and only optimize the AU's trajectory and BS's power allocation by using our proposed DDPG-based algorithm.
		\item \emph{EPA}: Furthermore, in this baseline, the users are assigned with equal power values by the BSs while our proposed algorithm solves the user association and trajectory optimization.     
\end{itemize}
\begin{figure}[t!]
	\centering
	\includegraphics[width=6.3cm, height=4cm]{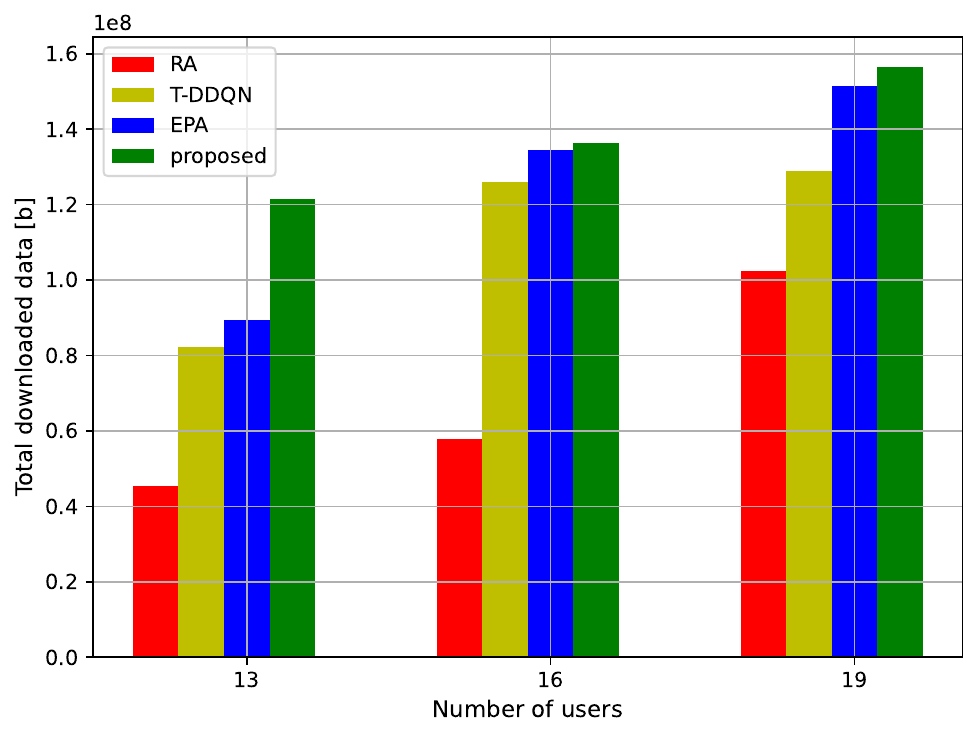}
	\caption{Total downloaded data with different number of users.}
	\label{fig:total_data}
\end{figure}

The optimal trajectories of AUs are illustrated in Fig. \ref{fig:opt_traj}, where we consider $4$ AUs in the considered area. We compare our proposed approach with T-DDQN in which the continuous action space is transformed into the discrete ones. As we can observe from Fig. \ref{fig:traj_proposed}, the AUs in our proposed method fly near the GBSs while traveling from their starting points to the destination points. As a result, the AUs can maintain better link quality with their associated GBSs or LEOSats. However, as shown in Fig. \ref{fig:traj_DDQN}, the AUs are flying near GBS-1 due to the accuracy loss of action space discretization. Therefore, it is difficult for the AUs to meet their QoS requirements. Moreover, their energy consumption will be extremely high to reach their destination points. This suggests that our proposed DDPG-based algorithm outperforms T-DDQN regarding the AUs' trajectories.

In Fig. \ref{fig:total_data}, we illustrate the total data service of the proposed network for different number of users. To show the superior performance of our proposed algorithm, we implement several baselines such as T-DDQN, RA, and EPA. As we can see from Fig. \ref{fig:total_data}, the total data service received by the users in RA is the minimum. This is because the users cannot establish a better link by the random association of BSs. On the other hand, the performance of EPA is better than that of T-DDQN. The reason is that our proposed DDPG-based trajectory optimization algorithm in EPA can determine the optimal values of two-dimensional coordinates and the altitudes of AUs to enhance the users' data service by properly reducing network interference as well as enhancing the link quality. Nevertheless, our proposed approach can yield the maximum data service for all the users in the proposed SAGIN. 

\begin{figure}[t!]
	\centering
	\includegraphics[width=6.3cm, height=4cm]{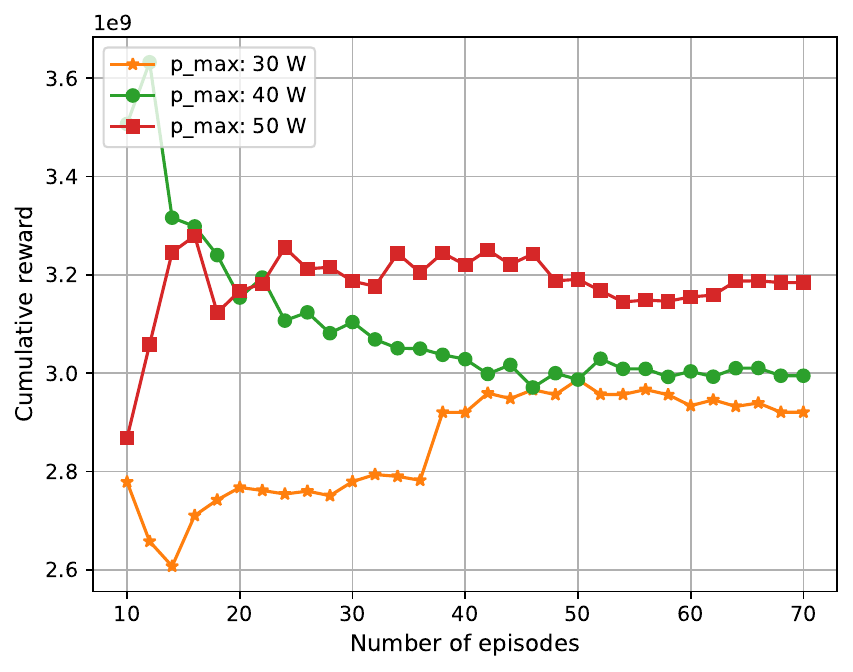}
	\caption{Cumulative reward for different power values.}
	\label{fig:reward}
\end{figure}

By considering the different power values of BSs, we evaluate the cumulative reward of the proposed data service maximization problem in Fig. \ref{fig:reward}. As presented in Fig. \ref{fig:reward}, the greater cumulative reward can be obtained with the increasing power values. This is because the higher data service can be obtained with the greater power value which in turn increases the reward as given in (\ref{reward_eq}). Moreover, we can observe that our proposed approach can guarantee the convergence of cumulative reward in all cases.

\section{Conclusion}  \label{conclusion}
In this paper, we have studied a SAGIN where the GBSs and LEOSats serve the coexisted AUs and GUs for downlink communication. Considering the spectrum scarcity, interference, and QoS requirements of the users, we have formulated the problem of maximizing the total downloaded data size of the users in the network by jointly optimizing the user association, AUs' trajectories, and power allocation of GBSs and LEOSats. Then, to address the non-convexity of the formulated mixed-integer non-convex optimization problem, we have divided it into two subproblems: 1) user association and 2) trajectory and power allocation problem. The user association problem is reformulated as a binary integer programming problem and solved by using the Gurobi optimizer, while the DDPG-based method is employed to address the trajectory and power allocation problem. After that, the two subproblems are iteratively solved until convergence is reached. Through extensive simulation results, we have verified that the performance of our proposed framework is $51.6$ \%, $19.2$ \%, and $10.4$ \% better than the baseline methods such as T-DDQN, RA, and EPA, respectively.

\bibliographystyle{IEEEtran}
\bibliography{letter_ref}

\begin{thebibliography}{10}
\providecommand{\url}[1]{#1}
\csname url@samestyle\endcsname
\providecommand{\newblock}{\relax}
\providecommand{\bibinfo}[2]{#2}
\providecommand{\BIBentrySTDinterwordspacing}{\spaceskip=0pt\relax}
\providecommand{\BIBentryALTinterwordstretchfactor}{4}
\providecommand{\BIBentryALTinterwordspacing}{\spaceskip=\fontdimen2\font plus
\BIBentryALTinterwordstretchfactor\fontdimen3\font minus \fontdimen4\font\relax}
\providecommand{\BIBforeignlanguage}[2]{{%
\expandafter\ifx\csname l@#1\endcsname\relax
\typeout{** WARNING: IEEEtran.bst: No hyphenation pattern has been}%
\typeout{** loaded for the language `#1'. Using the pattern for}%
\typeout{** the default language instead.}%
\else
\language=\csname l@#1\endcsname
\fi
#2}}
\providecommand{\BIBdecl}{\relax}
\BIBdecl

\bibitem{starlink}
``Direct to cell first text update,'' https://www.starlink.com/updates.

\bibitem{zhu2017non}
X.~Zhu, C.~Jiang, L.~Kuang, N.~Ge, and J.~Lu, ``Non-orthogonal multiple access based integrated terrestrial-satellite networks,'' \emph{IEEE Journal on Selected Areas in Communications}, vol.~35, no.~10, pp. 2253--2267, Oct. 2017.

\bibitem{fang2022noma}
X.~Fang, W.~Feng, Y.~Wang, Y.~Chen, N.~Ge, Z.~Ding, and H.~Zhu, ``{NOMA}-based hybrid satellite-{UAV}-terrestrial networks for 6{G} maritime coverage,'' \emph{IEEE Transactions on Wireless Communications}, vol.~22, no.~1, pp. 138--152, Jan. 2023.

\bibitem{lei2024spatial}
L.~Lei, A.~Wang, E.~Lagunas, X.~Hu, Z.~Zhang, Z.~Wei, and S.~Chatzinotas, ``Spatial-temporal resource optimization for uneven-traffic {LEO} satellite systems: Beam pattern selection and user scheduling,'' \emph{IEEE Journal on Selected Areas in Communications}, vol.~42, no.~5, pp. 1279--1291, May 2024.

\bibitem{lee2023interference}
H.-W. Lee, C.-C. Chen, S.~Liao, A.~Medles, D.~Lin, I.-K. Fu, and H.-Y. Wei, ``Interference mitigation for reverse spectrum sharing in {B5G/6G} satellite-terrestrial networks,'' \emph{IEEE Transactions on Vehicular Technology}, vol.~73, no.~3, pp. 4247--4263, Mar. 2024.

\bibitem{chen2022multi}
Q.~Chen, W.~Meng, T.~Q. Quek, and S.~Chen, ``Multi-tier hybrid offloading for computation-aware {IoT} applications in civil aircraft-augmented {SAGIN},'' \emph{IEEE Journal on Selected Areas in Communications}, vol.~41, no.~2, pp. 399--417, Feb. 2023.

\bibitem{choi2024latency}
J.~Choi, S.~Krishnan, and J.~Park, ``Latency-optimal resource allocation for {UAV}-aided {LEO} communication,'' \emph{IEEE Transactions on Vehicular Technology (Early Access)}, Mar. 2024.

\bibitem{tsai2024distributionally}
K.-C. Tsai, L.~Fan, R.~Lent, L.-C. Wang, and Z.~Han, ``Distributionally robust optimal routing for integrated satellite-terrestrial networks under uncertainty,'' \emph{IEEE Transactions on Communications (Earyl Access)}, May 2024.

\bibitem{pervez2021joint}
F.~Pervez, L.~Zhao, and C.~Yang, ``Joint user association, power optimization and trajectory control in an integrated satellite-aerial-terrestrial network,'' \emph{IEEE Transactions on Wireless Communications}, vol.~21, no.~5, pp. 3279--3290, May 2021.

\bibitem{mei2019joint}
H.~Mei, K.~Wang, D.~Zhou, and K.~Yang, ``Joint trajectory-task-cache optimization in {UAV}-enabled mobile edge networks for cyber-physical system,'' \emph{IEEE Access}, vol.~7, pp. 156\,476--156\,488, Oct. 2019.

\bibitem{jiang2022green}
X.~Jiang, M.~Sheng, Z.~Nan, X.~Chengwen, L.~Weidang, and W.~Xianbin, ``Green {UAV} communications for 6{G}: A survey,'' \emph{Chinese journal of aeronautics}, vol.~35, no.~9, pp. 19--34, Sep. 2022.

\end{thebibliography}
\end{document}